\documentclass[conference]{IEEEtran}
\IEEEoverridecommandlockouts
\usepackage{cite}
\usepackage{amsmath,amssymb,amsfonts}
\usepackage{algorithmic}
\usepackage{graphicx}
\usepackage{textcomp}
\usepackage{xcolor}
\def\BibTeX{{\rm B\kern-.05em{\sc i\kern-.025em b}\kern-.08em
    T\kern-.1667em\lower.7ex\hbox{E}\kern-.125emX}}
\begin{document}

\title{iASiS  Open  Data  Graph: Automated Semantic Integration of Disease-Specific Knowledge
\thanks{\textbf{\textcopyright2020 IEEE.}}
}

\author{\IEEEauthorblockN{Anastasios Nentidis\textsuperscript{1,2}, Konstantinos Bougiatiotis\textsuperscript{1,3}, Anastasia Krithara\textsuperscript{1} and Georgios Paliouras\textsuperscript{1}}
\IEEEauthorblockA{
\textsuperscript{1}\textit{Institute of Informatics and Telecommunications, NCSR Demokritos, Athens, Greece} \\
\textit{Email: \{tasosnent, bogas.ko, akrithara, paliourg\}@iit.demokritos.gr}\\
\textsuperscript{2}\textit{School of Informatics, Aristotle University of Thessaloniki, Thessaloniki, Greece} \\
\textsuperscript{3}\textit{Department of Informatics and Telecommunications, National and Kapodistrian University of Athens, Athens, Greece} \\
}}

\maketitle

\begin{abstract}
In biomedical research, unified access to up-to-date domain-specific knowledge is crucial, as such knowledge is continuously accumulated in scientific literature and structured resources.
Identifying and extracting specific information is a challenging task and computational analysis of knowledge bases can be valuable in this direction. However, for disease-specific analyses researchers often need to compile their own datasets, integrating knowledge from different resources, or reuse existing datasets, that can be out-of-date.
In this study, we propose a framework to automatically retrieve and integrate disease-specific knowledge into an up-to-date semantic graph, the \textit{iASiS Open Data Graph}. 
This disease-specific semantic graph provides access to knowledge relevant to specific concepts and their individual aspects, in the form of concept relations and attributes. 
The proposed approach is implemented as an open-source framework and applied to three diseases (Lung Cancer, Dementia, and Duchenne Muscular Dystrophy). 
Exemplary queries are presented, investigating the potential of this automatically generated semantic graph as a basis for retrieval and analysis of disease-specific knowledge.
\end{abstract}

\begin{IEEEkeywords}
biomedical knowledge, knowledge graphs, semantic integration, disease-specific, biomedical literature
\end{IEEEkeywords}
\section{Introduction}\label{sec:intro}

A lot of biomedical knowledge is published every day in the literature and structured resources like biomedical ontologies. It is a challenge for biomedical experts to identify and process all available knowledge. For example, 1.3 million citations were added to MEDLINE/PubMed during 2018\footnote{
https://www.nlm.nih.gov/bsd/licensee/baselinestats.html}, which corresponds to more than two citations per minute.
In this context, the identification of articles relevant to a specific research topic can be challenging.
Efficient access to relevant knowledge is crucial and simple term-based search can retrieve irrelevant documents (e.g. due to homonyms) or miss relevant documents (e.g. due to synonyms, abbreviations or term mismatch). 
Much effort has been made to address this issue, including semantic search approaches that use predefined concepts which can have several associated synonyms and relations with other concepts, expanding the query terms. 

PubMed\footnote{https://www.ncbi.nlm.nih.gov/pubmed/} is an established knowledge resource considered for biomedical literature and forms the basis for a variety of search tools \cite{Lu2011}.
PubMed supports semantic search based on the Medical Subject Headings (MeSH) hierarchy\footnote{https://www.nlm.nih.gov/mesh/}. A team of curators in the U.S. National Library of Medicine (NLM) continuously annotates articles added in PubMed with the appropriate MeSH terms that represent the topics of each article.
These topic annotations can be exploited for information retrieval and knowledge extraction in the form of relations between MeSH terms, as overviewed by Zhang \textit{et al.} \cite{Zhang2014}.

Gathering a set of articles relevant to a topic of interest is often not sufficient. An article may contain different pieces of knowledge, which are more or less relevant to the interests of a researcher. Additionally, the value of these knowledge items can change, if they are combined with information from other articles or resources. A knowledge base supports this process of organizing and storing domain knowledge, in order to be easily accessible both for users, through adequate interfaces, and for computational analysis.
Biomedical knowledge bases, such as DrugBank \cite{Wishart2008},
though manually curated by domain experts, are usually supported by computational tools that analyze the literature.

A variety of text-mining approaches has been proposed for the automated extraction of knowledge from literature, including term co-occurrence and syntactic analysis of article text \cite{Rebholz-Schuhmann2012}. Such approaches usually extract binary relations between biomedical entities such as interactions between proteins or gene-disease associations.
Different biomedical knowledge bases have been automatically developed, based on such approaches, integrating knowledge extracted from various textual resources such as KnowLife \cite{Ernst2015}. The Semantic MEDLINE\cite{Rindflesch2017} is one of the most systematic approaches, providing uniform access to knowledge from MEDLINE abstracts matching a query visualized as an interactive network of concepts linked by a range of relations.

On the other hand, integration of structured knowledge from different biomedical ontologies and databsases is another challenge, and as in other fields, semantic web technologies have been exploited in this direction. 
In KaBOB \cite{Livingston2015} for example data from 14 ontologies and 18 databases have been integrated into a common RDF-based knowledge base. 
Another recent effort to biomedical knowledge integration is the Hetionet\cite{himmelstein2017systematic} which exploits the power of graph databases to develop an heterogeneous network integrating knowledge from a variety of resources. This work focuses on structured resources, incorporating a limited number of relations extracted from selected MEDLINE articles through co-occurrence analysis. 

Though many different knowledge bases have been developed in the field of biomedicine, we believe that an open-source solution for the automated development of an up-to-date and comprehensive disease-specific knowledge graph, from both literature-extracted and structured knowledge, is still missing. Therefore, we present the \textit{iASiS Open Data Graph} framework as a flexible pipeline of independent software modules for the construction and maintenance of a disease-specific open knowledge graph for any disease of interest.

The rest of this article is structured as follows. In Section~\ref{sec:meth} we introduce the proposed \textit{iASiS Open Data Graph} framework and describe its components. In Section~\ref{sec:cases} we apply the framework to create distinct knowledge bases for three diseases and present some example queries to illustrate and discuss the properties of the resulting knowledge graphs. Finally, in Section~\ref{sec:conc} we draw some conclusions based on the reported experiments.



\section{Methods}\label{sec:meth}
The \textit{iASiS Open Data Graph} provides a framework to combine state-of-the-art tools to automatically retrieve, extract and integrate knowledge from open structured resources and disease-specific literature into a semantically integrated knowledge graph.
In this framework, instead of adopting an RDF-based semantic schema, we opt for semantic integration based on the Unified Medical Language System (UMLS) \cite{bodenreider2004unified}, which is structured and comprehensive enough to express a wide range of biomedical knowledge in a standard way, but still quite close to natural language and intuitive to be directly presented to human users. 

In particular, user interfaces such as Neo4j Bloom\footnote{https://neo4j.com/bloom/} can be used for interactive visualization of the knowledge so that domain experts can access and edit it.
On the other hand, the knowledge can be accessed computationally, though graph queries, providing up-to-date datasets for a variety of knowledge-analysis studies and applications. For example, datasets  developed through the \textit{iASiS Open Data Graph} have already been used in a knowledge-driven framework for supporting personalized medicine \cite{Vidal2019}, in a knowledge-graph-based method for prediction of drug-to-drug interactions \cite{Aisopos2019} and in a path-based method for detecting erroneous edges in knowledge graphs \cite{Fasoulis2020}.

\subsection{Framework architecture}
\begin{figure}[tbp]
\centerline{\includegraphics[width=.48\textwidth]{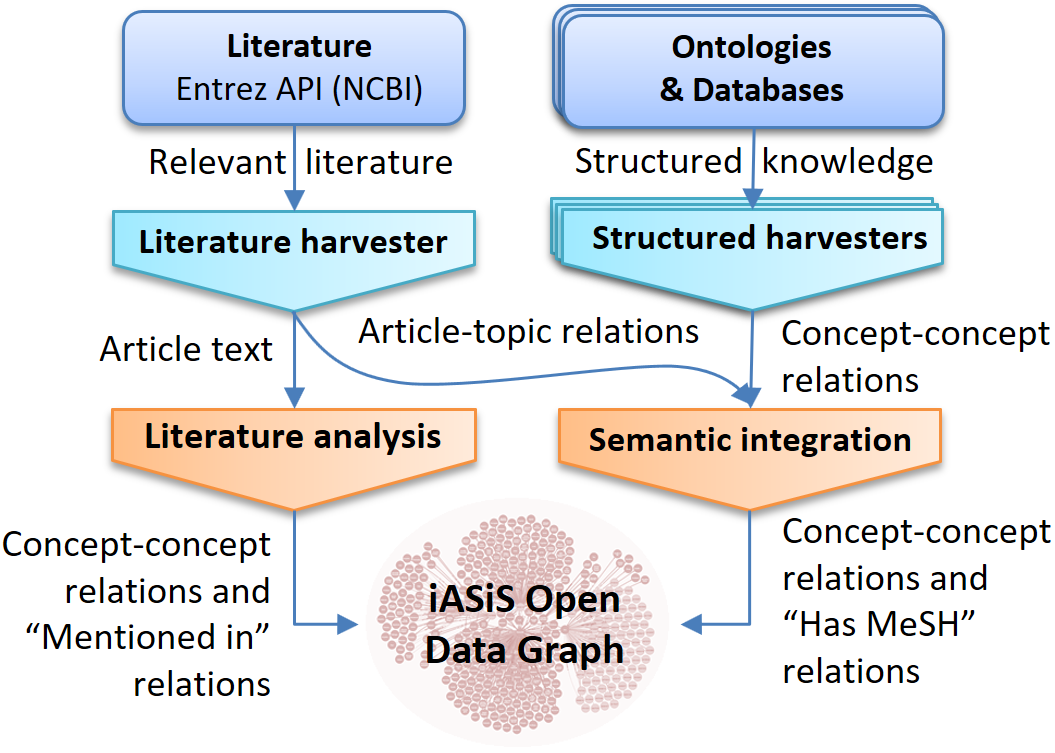}}
\caption{The architecture of the the iASiS Open Data Graph framework.}\label{fig:00}
\end{figure}

The proposed approach for semantic retrieval, extraction and integration of disease-specific knowledge was designed and developed as a framework of distinct modules that perform well-defined tasks and can be reused independently. The architecture of the framework is presented in  Figure~\ref{fig:00}.
At first, the \textit{Literature harvester} module interacts with the Entrez API of the NCBI\footnote{https://www.ncbi.nlm.nih.gov/books/NBK25497/} for the online semantic retrieval of all literature available for the disease of interest.
Then, the \textit{Literature analysis} module employs state-of-the-art tools on biomedical natural language text analysis to extract structured knowledge, in the form of a graph of inter-related concepts linked to the resource documents. 
In addition, the \textit{Structured harvester} modules extract binary relations between concepts from biomedical structured resources and the \textit{Semantic integration} module maps all the entities in the UMLS coding system. 
Finally, all the knowledge is integrated in a graph database, under a simple but powerful representation, where graph queries can be employed to serve the information needs of disease-specific biomedical research.

The same modules can also be employed for the automated update of the knowledge graph. Given the date of the last update, the \textit{Literature harvester} can retrieve new relevant articles and the \textit{Literature analysis} modlue can extract new knowledge from them and update the graph. 
The result is a semantically integrated disease-specific graph with up-to-date automatically extracted knowledge from literature and high-quality reviewed knowledge from ontologies and databases. 

\subsection{Data harvesting}\label{DH}

The scientific literature is a basic and up-to-date resource of biomedical knowledge. 
PubMed offers more than 28 million citations and is uniformly accessible, through the Entrez REST API, with PubMed Central\footnote{https://www.ncbi.nlm.nih.gov/pmc/} (PMC) which offers the full-text of about 4.7 million articles. 
In this work we exploit this API, developing the \textit{Literature harvester} module to search in PubMed based on the MeSH descriptor of a disease, in order to identify the relevant articles available. Then, it retrieves all the documents identified and extracts the abstract text and the topic annotations, out of the more than 100 different types of hierarchically related elements available in the MEDLINE/Pubmed XML format. In addition, it also retrieves corresponding records from PMC, when available, and extracts the full-text body from the PMC XML format.
The inclusion of full-text in this framework is important to perform an integration of knowledge as deep as possible, including specific details available only in full texts. 

On the other hand, biomedical ontologies are an important source of manually-curated and well-structured domain knowledge and more than 140 biomedical ontologies are available from the  Open Biomedical Ontologies (OBO) Foundry\footnote{http://www.obofoundry.org/}.
For this reason, a \textit{Structured harvester} was developed to extract hypernymic relations (e.g. \textit{is-a}) between concepts from ontologies in the OBO format for integration in the \textit{iASiS Open Data Graph}. Although other types of relations are provided by some ontologies, we decided to focus on hypernymic relations as the most important and universal knowledge type.  

Apart from OBO ontologies, knowledge from other structured resources can also be integrated, as long as it can be expressed as relations between concepts mapped to the UMLS. To do so, each structured resource should either be transformed into the OBO format, or a specific \textit{Structured harvester} should be developed to extract the knowledge as relations between concepts. In this work, we developed the \textit{Structured harvesters} focusing on the most basic and abundant relation types for each resource. However, extending the harvesting modules to extract more relation types can be supported by the framework.  
For example, in order to support the use cases presented in Section-\ref{sec:cases} we developed a pre-processing script to convert the original MeSH XML file into a simple OBO-based version of MeSH, and a \textit{Structured harvester} to extract drug-to-drug interactions from the original XML file of DrugBank\cite{wishart2017drugbank}. 

All \textit{Structured harvesters} produce datasets of relations in a common JSON format.
The same stands for topic annotations from the \textit{Literature harvester}, as relations between articles and MeSH descriptors. 
Thus, all harvested structured data are handled by the same \textit{Semantic integration} module to be mapped under the UMLS Schema, if needed.


\subsection{Literature analysis}\label{LA}

Text mining offers the potential to tap into the knowledge still buried in the ever-increasing body of biomedical literature. 
Our goal is not to create a new text mining tool, but rather to create a framework where any such tool can be incorporated to extract biomedical entities and relations from text for enriching a unified knowledge graph.
Working towards this goal, we need to recognize biomedical entities in text and also map them to a common semantic schema. 
Though a variety of specialized biomedical terminological resources is being developed, such as ontologies and lexicons, they are usually non inter-operable as they rely on different coding systems.

In this work, we adopt the UMLS Metathesaurus \cite{schuyler1993umls} as the reference schema for semantic integration of entities.
The key aspect of this thesaurus is that differing names and identifiers for a biomedical entity in different vocabularies are linked under a single UMLS \textit{concept}. Therefore, the Metathesaurus deals with term variation and at the same time creates links between different vocabularies. 
Currently, it contains more than than 100 vocabularies and more than one million concepts, accumulating knowledge from different domains (e.g. chemical, phenotypic), and is also regularly expanded with new resources. 


In addition, we also adopt the UMLS Semantic Network (SN)\cite{mccray2003upper} which enriches the concepts of the Metathesaurus with semantic types and defines types of semantic relations between them. 
There are 133 hierarchically structured semantic types expressing a high-order grouping of the concepts into categories, such as diseases, genes or genomes etc. 
Together with 55 semantic relations between these types, a rich semantic network is defined, spanning the whole biomedical domain.


In order to harness the power of this UMLS-based representation, we use SemRep \cite{rindflesch2003interaction}, a tool that extracts predications between biomedical entities, i.e. semantic triples in the form of subject-predicate-object, from unstructured text. 
Each entity is a UMLS concept and each predicate is a relation of the semantic network, connecting the semantic types of these two concepts in the context of the specific sentence. Each predication, is also annotated by SemRep with a value indicating whether a negation was recognized in the text for this predication.
For concept extraction, SemRep relies on MetaMap \cite{aronson2001effective} which is a tool that uses symbolic natural-language processing (NLP) and computational-linguistic techniques to map biomedical text to Metathesaurus concepts. 
Evaluation in certain predicates suggests that SemRep predications are quite precise, with precision ranging from 75\% to 96\%, but can miss in recall, which ranges between 55\% and 70\%~\cite{Kilicoglu2012}.

\begin{figure}[tbp]
\centerline{\includegraphics[width=.45\textwidth]{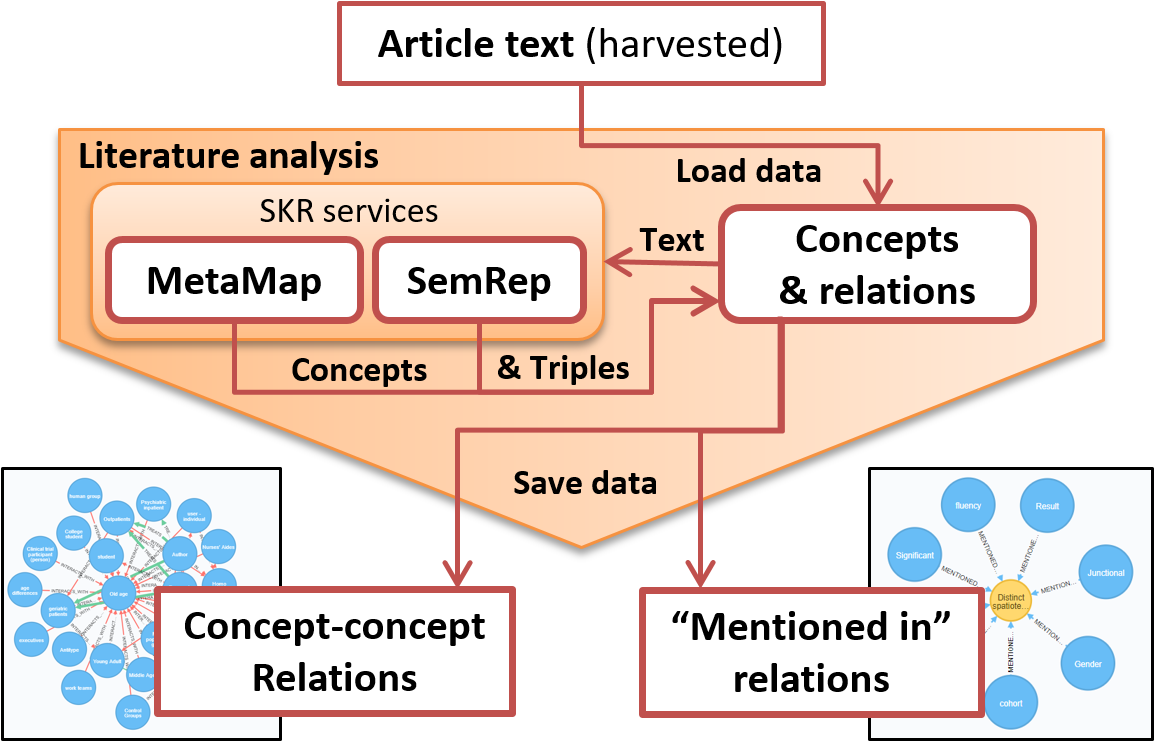}}
\caption{The literature analysis module.}\label{fig:02}
\end{figure}

For the \textit{iASiS Open Data Graph} framework, we developed a \textit{Literature analysis} module that uses these tools to analyze the harvested literature as shown in Figure~\ref{fig:02}.
Specifically, after some preprocessing (e.g. removal of tables, LaTeX code etc.) SemRep and MetaMap process the abstract and the full text of the articles to extract concepts and predications\footnote{SemRep V1.7, MetaMap 2016V2, UMLS Metathesaurus 2015AA.}. 
The module then transforms the recognized entities and triples into corresponding nodes and edges in a graph. 
Alongside the extracted concept-to-concept predications stemming from the Semantic Network, we also add in the same graph edges between nodes of articles and nodes of concepts, that denote the occurrence of a specific concept in the corresponding article. We call this new type of relation ``\textit{Mentioned in}''. 
The motivation behind this approach is to integrate information required for co-occurrence analysis, while maintaining the provenance of the knowledge extracted though syntactic analysis by SemRep. 
Co-occurrence analysis can provide high-level associations between two concepts and help leveraging knowledge from multiple articles. 


\subsection{Knowledge Graph}
The \textit{iASiS Open Data Graph} integrates both structured and unstructured knowledge in the same semantic graph. To accomplish that, we use a single node per UMLS concept or MEDLINE/PubMed article. 
A ``\textit{Mentioned in}'' edge connects each concept with each article it has been extracted from. 
In addition, a new type of article-to-concept relation, which we call ``\textit{Has MeSH}'', was added to integrate the manual MeSH topic annotations harvested from PubMed. 
In order to integrate such entities from structured sources in the same graph, we took advantage of the UMLS REST API\footnote{https://documentation.uts.nlm.nih.gov/rest/home.html}, which provides mappings from different vocabulary identifiers (e.g. MeSH, DrugBank, GO) to UMLS concepts. 
For example, for each UMLS concept corresponding to any MeSH descriptors of an article the \textit{Semantic integration} module adds in the graph a triple in the form of ``article-Has MeSH-concept''.  


Using a single node per concept leads to a highly integrated graph, where the knowledge about an entity is represented in the form of interactions with other concepts and articles. The actual sources of this knowledge, such as ontologies or specific articles, are stored as provenance properties of the links.
Regarding the technical implementation of the knowledge graph  we used the community edition of the  Neo4j\footnote{https://neo4j.com/}, a graph database which is increasingly adopted for bioinformatics projects that model biological connectivity\cite{yoon2017use, himmelstein2017systematic}.
Graph databases are a natural choice for graph storage as explicitly focus on the connectivity between the nodes and have performance superior to traditional SQL databases, when traversing many levels of connectivity \cite{jaiswal2013comparative}.


\section{Case studies and discussion}
\label{sec:cases}
The \textit{iASiS Open Data Graph} framework was applied to two prevalent and one rare disease, namely Lung Cancer (LC), Dementia including Alzheimer Disease (ADD) and Duchenne Muscular Dystrophy (DMD). Therefore, a disease-specific knowledge graph was developed for each use case. Exemplar graph queries were used to extract and compare knowledge from these graphs, investigating their potential. 
The source code of the framework is openly available on GitHub\footnote{https://github.com/tasosnent/Biomedical-Knowledge-Integration}, as well as all the graph queries used for the tables and the examples presented in this section. 

\subsection{Dataset creation}
The framework was employed once for each disease, configured with the appropriate semantic topic\footnote{MeSH topics ``Lung Neoplasms'', ``Dementia'' and ``Muscular Dystrophy, Duchenne'' 
for LC, ADD and DMD respectively.} to retrieve and analyse all available relevant literature in PubMed and create the corresponding semantic graphs. 
As expected, the volume of available knowledge was different for the three cases resulting in graphs of different sizes. For ADD and LC which are highly prevalent diseases, more than 100,000 articles were available. However, even for DMD, which is a rare disease, the number of directly relevant articles exceeded 4,000.

As shown in Table~\ref{Tab:01}, for all three use cases only as small portion of the relevant articles had the full text available.
DMD had the higher percentage of full-text availability exceeding 24\%, probably because the relevant literature has been published more recently. In terms of quantity, concept occurrence relations (``\textit{Mentioned in}'') is dominant and topic annotations (``\textit{Has MeSH}'') are also quite abundant, while domain knowledge, in the form of UMLS SN relations between extracted concepts is the least frequent. This could be attributed to limitations in relation extraction step which is the most complex and less mature part of the processing.  

\begin{table}[tbp]
\caption{ Disease-specific literature and knowledge extracted from it. 
\label{Tab:01}} 
{\begin{tabular*}{\linewidth}{@{}llll@{}}
\textbf{Case study}                      & \textbf{DMD}       & \textbf{ADD}       &  \textbf{LC}        \\ 
Articles                        & 4,403     & 108,458   & 141,712   \\
Articles with full-text         & 1,075	    & 6,000	    & 10,000      \\
UMLS concepts extracted         & 21,982   & 75,985    & 92,846    \\
UMLS SN relations extracted     & 27,954    & 392,421   & 608,759   \\
Has-MeSH relations              & 113,136   & 3,651,698 & 4,228,785 \\
Mentioned-in relations          & 335,071   & 7,587,772 & 9,940,847 \\
\end{tabular*}}{
}
\end{table}


\begin{table}[tbp]
\caption{ Knowledge integrated from structured resources.
\label{Tab:02}} 
{\begin{tabular*}{\linewidth}{@{}llll@{}}
\textbf{Resource} & \textbf{UMLS Concepts} & \textbf{Relation type} & \textbf{Relations} \\
DO  & 5,307 & is a & 5,129\\
GO  & 64,751 & is a & 125,629 \\
MeSH & 55,400 & is a & 123,287\\
DrugBank & 3,642 & drug interactions & 1,628,077\\
\end{tabular*}}{ }
\end{table}

As described in Section \ref{sec:meth}, knowledge from structured resources can also be integrated in the graph. 
These structured resources considered here are not disease-specific and details about the corresponding datasets are presented separately in Table~\ref{Tab:02}.
In particular, two basic ontologies have been selected for these experiments.
Namely, the Gene Ontology (GO) \cite{ashburner2000gene}, that provides more than 24,000 concepts to represent basic categories of genomic knowledge, and the Disease Ontology (DO) \cite{Schriml2012}, that semantically integrates more than 10,000 concepts from different resources, with more than 5,000 mappings to the UMLS\footnote{Only concepts with UMLS mappings are integrated in this framework}.
Another important resource of hypernymic relations is MeSH, which provides more than 28,000 hierarchically organized topical descriptors used for semantic indexing in PubMed.
Finally, we also integrate drug-to-drug interactions from DrugBank, which is a comprehensive, manually maintained resource with more than 10,000 drug entries
and more than 200 data fields per drug. Drug interactions, which are the most abundant information in DrugBank exceeding 300,000, are of great interest when studying diseases \cite{wishart2017drugbank}. 

 In this setup, only hypernymic relations and drug-to-drug interactions were harvested from structured resources. These types of relation are among the most important ones and constitute a proof of concept for the integration of any kind of relation from structured resources. 
 In practice, different setups combining relevant resources or parts of them would be more suitable for each new use case. Apart from selecting relevant resources other important issues, like the potential overlap or conflict in their content, should also be carefully considered.

\subsection{Query examples}


The amount of accumulated knowledge and the variability observed among different diseases highlight the importance of two distinct but complementary needs regarding knowledge access. It is crucial to have precise access to highly detailed information and at the same time have a broad overview of all knowledge available to select areas to focus. 
The integrated disease-specific semantic graph produced by the \textit{iASiS Open Data Graph} framework can support both these needs. 

For example, ranking the semantic types by the number of distinct concepts extracted from literature for each disease can indicate directions for further study. 
Some semantic types, are ranked high for all three diseases, such as \textit{Gene or Genome}\footnote{\textit{Gene or Genome} is ranked first in LC and ADD, and second in DMD.}.
On the other hand, some interesting differences can be observed for other semantic types, such as \textit{Neoplastic Process} which is ranked higher for LC 
as can be seen in Table~\ref{Tab:04}. 
\begin{table}[tbp]
\caption{ 
Top 10 semantic types with the highest standard deviation (STDV) of ranks for the three use cases.
\label{Tab:04}} {
\begin{tabular*}{\linewidth}{@{}lcccl@{}}
\textbf{Semantic type } & \multicolumn{3}{c}{\textbf{Disease rank}}  & \textbf{STDV}     \\
  & \textbf{LC}  & \textbf{ADD}  & \textbf{DMD}  &      \\
Plant & 23      & 22      & 79       & 32.62 \\
Bacterium& 53      & 72      & 104      & 25.78 \\
Neoplastic Process& 9       & 36      & 59       & 25.03 \\
Fungus& 72      & 94      & 120      & 24.03 \\
Mental or Behavioral Dysf. & 65      & 28      & 53       & 18.88 \\
Eukaryote& 39      & 43      & 70       & 16.86 \\
Activity& 81      & 76      & 51       & 16.07 \\
Mental Process& 62      & 39      & 34       & 14.93 \\
Hazardous or Poisonous Subst.& 32      & 45      & 61       & 14.53 \\
Organism Attribute& 77      & 71      & 50       & 14.18 \\
\end{tabular*}}{ }
\end{table}
In order to emphasize the differences among diseases, semantic types in Table~\ref{Tab:04} are ordered by the decreasing standard deviation of their ranks between the three diseases. It is interesting that the semantic type \textit{Plant} has the most differing rankings for the three diseases and is more frequent in LC and ADD than in DMD.
Such observations can provide directions for further study, constructing a profile for each disease.

Next, we will focus on the previously observed importance of Plants in LC and ADD,
querying for the five most frequent concepts with Semantic Type \textit{Plant}\footnote{ The five most frequent \textit{Plant} concepts are [Plants; Nicotiana; Gossypium; Bikinia le-testui; Rosa] for LC and [Bark - plant part; Parkinsonia; Bikinia le-testui; Plants; Ginkgo biloba] for AD.} in the corresponding graphs. 
Examination of some source articles of occurrence reveals that some of them are indeed plants of interest for the research on LC (\textit{Nicotiana}, \textit{Gossypium}) and ADD (\textit{Ginkgo biloba}) respectively. The  fact  that  articles  are  organized per occurring concept allows for a selective examination of them.
In general, a quick overview of the knowledge available for each concept can be retrieved with corresponding graph queries.

In the LC graph, for example, the concept \textit{Plants} occurs 1,463 times in 795 articles and 13 articles have \textit{Plants} as a topic. In addition, there are 195 distinct relations of 5 types (``location of'', ``is a'', ``process of'', ``part of'', ``interacts with'') between \textit{Plants} and 194 distinct concepts occurring in the LC literature. 
An examination of the articles for the ten concepts more frequently related with \textit{Plants}\footnote{Curcuma longa; Chrysotile; polyphenols; 3-hydroxyflavone; Asbestos; Antineoplastic Agents; Oils, Volatile; Chlorophyll; Magnolia}
confirms that most of them are plant species (e.g. \textit{Curcuma longa}, \textit{Magnolia}) or chemicals found in plants (e.g. \textit{polyphenols}, \textit{3-hydroxyflavone}, \textit{Chlorophyll}) that have been studied for potential effect on LC. 

\begin{table}[tbp]
\caption{Drugs related to \textit{Long Term Survivorship} (LTS) in LC literature enriched with other interacting concepts.
\label{Tab:08}} {
\begin{tabular*}{\linewidth}{@{}lll@{}}
\textbf{Concept label}                   & \textbf{Interacting concepts} &  \textbf{Interacting enzymes} \\
Cisplatin                       & 991 (538*)  &  45  \\
Antineoplastic Agents           & 210   &  24 \\
Aim                             & 243   &  11  \\
Melphalan                       & 58 (51*)   &  0 \\
everolimus                      & 640 (627*)   &  2  \\
cetuximab                       & 71 (31*)   &  4  \\
C3a des-Arg77                   & 6     &  0  \\
Interferons                     & 12    &  0  \\
animal allergen extracts        & 87    &  5  \\
gefitinib                       & 985 (760*)   &  39  \\
Altretamine                     & 110 (101*)  &  0   \\
Topotecan                       & 561 (532*)  &  2  \\
Paclitaxel                      & 1,518 (1,356*)  &  17 (6*) \\
Carboplatin                     & 147 (97*)  &  1 \\
\multicolumn{3}{l}{* Distinct concepts with interactions from DrugBank.}
\end{tabular*}}{}
\end{table}

In an alternative scenario, a researcher may be interested in the effect of drug combinations in long surviving LC patients.
A central concept in this case is the \textit{Long Term Survivorship} (LTS) of patients. This concept is mentioned in 2303 articles in the LC dataset, but none of them is annotated with it as a topic. 
We further segment this set of articles identifying more than 300 concepts co-occurring with LTS, that are also children of the \textit{Pharmaceutical Preparations} concept (i.e. related with ``is a'').
Apart from co-occurrence, we also identify 14 distinct concepts directly related with LTS, through five distinct types of relation, which are presented in Table~\ref{Tab:08}. 

Each concept can also be directly enriched with supplementary information such as, the number of distinct biomedical concepts interacting with it, or interacting enzymes in particular. 
In this example the relations extracted from the literature are not distinguished from relations from structured resources. In particular, ``is a'' relations from both literature and ontologies were used to retrieve drug concepts and ``Interacts with'' relations from both literature and DrugBank were considered for interacting entities in Table~\ref{Tab:08}. However, we could restrict into interactions from DrugBank which is manually cureated, to increase precision to the detriment of recall, as shown in Table~\ref{Tab:08}. Interactions extracted from literature that are not confirmed by DrugBank, should be considered more cautiously.


Finally, we can also exploit the graph structure of the knowledge to retrieve paths between entities in a disease-specific context. For example, the shortest paths between LTS and \textit{Drug Combinations} consist of two hops and an intermediate node. In some cases, this node is a relevant article connected with both concepts of interest through ``\textit{Mentioned in}'' and ``\textit{Has MeSH}'' edges. In other cases, this node is a chemical (such as \textit{Carboplatin} and \textit{Topotecan}) that ``co-exist with'' or ``interact with'' \textit{Drug Combinations} and can also ``affect'' LTS. As a next step, all the paths connecting the two concepts could also be considered. Though long paths can be too noisy to be directly handled by humans, they can be useful for computational analyses. For example, we can aggregate them to produce feature representations of concept pairs for training predictive models, as done for predicting pairs of interacting drugs in the iASiS\footnote{http://project-iasis.eu/} project \cite{Aisopos2019}.

\section{Conclusion}\label{sec:conc}
In this work we propose an open-source framework for the retrieval and semantic integration of disease-specific knowledge, focusing on automation and incremental update. Knowledge is extracted from relevant publications and integrated with knowledge from structured resources into a common semantic graph. 
This graph can can be queried by biomedical experts to access disease-specific domain knowledge in a uniform way, while also providing up-to-date datasets for applying and developing knowledge discovery methods.

The \textit{iASiS Open Data Graph} has been used to create semantic graphs for three diseases and example queries have been employed to investigate the potential and the limitations of the framework. The modular architecture of the framework allows the selective reuse of any of its components. 
Our future plans include the extension of the framework to harvest more relation types and structured resources and comparison of the integrated graph with background knowledge for automated noise removal. A detailed evaluation is planned about the usefulness of the produced disease-specific semantic graphs and the importance of observed and new caveats, in order to set priorities for future work.




\section*{Funding}

This work was supported by the EU H2020 programme, under grant agreement No 727658 (project iASiS).

\bibliographystyle{unsrt}
\bibliography{main}

\end{document}